# Electric-field-induced phase transition of <001> oriented Pb(Mg$_{1/3}$Nb$_{2/3}$)O$_3$-PbTiO$_3$ single crystals

Ke-Pi Chen[1], Xiao-Wen Zhang[1] and Hao-Su Luo[2]

[1]State Key Lab of New Ceramics and Fine Processing, Department of Materials Science and Engineering, Tsinghua University, Beijing 100084, P. R. China

[2]State Key Lab of High Performance Ceramics and Superfine Microstructure, Shanghai Institute of Ceramics, Chinese Academy of Sciences, Shanghai 201800, P. R. China

E-mail: chenkepi99@mails.tsighua.edu.cn

Abstract:

<001> oriented 0.7Pb(Mg$_{1/3}$Nb$_{2/3}$)O$_3$-0.3PbTiO$_3$ single crystals were poled under different electric fields, i.e. E$_{poling}$=4 kV/cm and E$_{poling}$=13 kV/cm. In addition to the temperature-dependent dielectric constant measurement, X-ray diffraction was also used to identify the poling-induced phase transitions. Results showed that the phase transition significantly depends on the poling intensity. A weaker field (E$_{poling}$=4 kV/cm) can overcome the effect of random internal field to perform the phase transition from rhombohedral ferroelectric state with short range ordering (microdomain) FE$_{SRO}$ to rhombohedral ferroelectric state with long range ordering (macrodomain) FE$_{lRO}$. But the rhombohedral ferroelectric to tetragonal ferroelectric phase transition originating from <111> to <001> polarization rotation can only be induced by a stronger field (E$_{poling}$=13 kV/cm). The sample poled at E$_{poling}$=4 kV/cm showed higher piezoelectric constant, $d_{33}$>1500 pC/N, than the sample poled at E$_{poling}$=13 kV/cm.

It was observed about twenty years ago that $Pb(Zn_{1/3}Nb_{2/3})O_3$-$PbTiO_3$(PZN-PT) single crystals with compositions close to the morphotropic phase boundary (MPB) exhibited anomalously large piezoelectric and electro-mechanical coupling constants, particularly when the samples were poled along <001> axes [1,2]. But it didn't attract much attention until several stimulating papers appeared in 1997 [3-6]. Recently both theoretical and experimental results provided a better understanding to the abnormal properties of single crystal PZN-PT and $Pb(Mg_{1/3}Nb_{2/3})O_3$-$PbTiO_3$ (PMN-PT) in the MPB region. Fu and Cohen [7] pointed out that a large piezoelectric response could be driven by polarization rotation induced by an external electric field in these single crystals. Almost at the same time, Noheda et al [8-10] used synchrotron x-ray powder diffraction measurements to reveal the existence of a monoclinic phase between the previously established tetragonal and rhombohedral regions in $Pb(Zr,Ti)O_3$ (PZT) which further supports the concept of polarization rotation. Subsequently, similar experiments were carried out on PZN-PT and PMN-PT single crystals [11-14], and it was found that monoclinic $FE_M$ or orthorhombic $FE_O$ phases can also be induced in a narrow range of compositions with proper poling history. But the rhombohedral phase $FE_R$ may also rotate to tetragonal phase $FE_T$ under an external electric field via an alternative path, which is different from that in PZT.

Only a few published papers have focused on the phase transition behavior and piezoelectric response of <001> poled PMN-PT single crystals with composition close to the MPB. By measuring the temperature dependence of the dielectric constant of a poled <001> oriented 0.7PMN-0.3PT single crystal, Viehland et al. [15] suggested that the transition from a normal ferroelectric state (macro domain) to a relaxor state (micro domain) with increasing temperature is responsible for the secondary transformation near 90 °C. For a poled <001> oriented 0.67PMN-0.33PT single crystal sample, Lu et al [16] also found an additional ferroelectric phase transition (FE-FE) at 80 °C using dielectric constant measurements. But they were uncertain whether this phase change corresponds to a $FE_O$-$FE_T$ or a $FE_R$-$FE_O$.

In this letter it is shown that the phase transition of poled <001> oriented 0.7PMN-0.3PT single crystals depends significantly on the intensity of the poling field. Dielectric constant measurements indicate that in addition to a transition at $T_{max}$=147 °C, there is a second phase transition far below $T_{max}$. X-ray diffraction experiments reveal that the crystalline state of the field-induced transition depends on the poling history of the crystal.

Single crystals of 0.7PMN-0.3PT were grown by the modified Bridgman method described in ref. [17]. It was reported that the segregation behavior during crystal growth results in compositional inhomogeneities of the PMN-PT single crystals even in the same boule[17]. Thus, the composition of the specimen was identified using X-ray fluorescence analysis to make sure it is 0.7PMN-0.3PT. In this study, only one <001> oriented 5x10x0.5 mm$^3$ specimen was used in all experiments to keep the composition unchanged. The sample was electroded with silver paste and poled at room temperature under various electrical fields (1st cycle: 4 kV/cm, 2nd cycle: 13 kV/cm). The dielectric constant was measured as a function of temperature at a heating rate of 4 °C/min using an HP4192A Precision LCR meter at different frequencies (0.1, 1, and 10kHz). XRD experiments were performed on Rigaku D/max-3B diffractometer using Cu-K$_\beta$ monochromatic radiation ($\lambda$=1.39223Å). Piezoelectric constant ($d_{33}$) of the samples with different poling field were measured using a Berlincourt-type quasistatic $d_{33}$ meter.

Figs. 1a-1c show the dielectric constants plotted as a function of temperature for unpoled (Fig.1a) and poled <001> oriented samples (Fig.1b, E$_{poling}$=4 kV/cm and Fig.1c, E$_{poling}$=13 kV/cm) at different frequencies under zero field heating (ZFH) respectively. For the unpoled sample, only one phase transition occurred at 147 $^o$C and the dielectric constant $\varepsilon$ showed weak frequency dependence. It can be attributed to a phase transition between a cubic state and a rhombohedral ferroelectric state with short range ordering (microdomain) FE$_{SRO}$. In contract with the unpoled sample, both samples of E$_{poling}$=4 kV/cm and 13 kV/cm demonstrate field-induced secondary transition around 90 $^o$C as shown in Fig.1b and Fig.1c.

It is difficult to distinguish the difference between the secondary transitions in Fig. 1b and Fig. 1c by dielectric constant measurements. But the XRD patterns of these two samples are quite different. Fig. 2a shows the (004) diffraction line (2$\theta$=87.50$^o$) of the unpoled sample. This symmetric profile has a relative wide FWHM (full width of half maximum) angle of $\Delta 2\theta$=0.56$^o$, which implies that a pseudocubic phase FE$_{SRO}$ exists in the unpoled 0.7PMN-0.3PT single crystals. This is in consistent with the results of the dielectric measurement. According to the Scherrer equation [18], the mean dimension of microdomain can be estimated by using the broadening of diffraction line. To make a comparision between Fig.2a and fig. 2b, it showed that the mean dimension of microdomain of unpoled sample is around 10nm. The lattice parameter $a$ of this form was calculated from (400) line. It equals to 0.402nm, which is almost identical with the results of Singh et al [19] ($a$=o.4023nm) and Zhang et al [20] ($a$=0.4019nm) for the 0.7PMN-0.3PT ceramics with pseudo-cubic form too. The existence of rhombohedral ferroelectric state with short range ordering FE$_{SRO}$ is in agreement with the results shown in [15] for the unpoled single crystals with same composition and can be explained by the effect of the internal random field [21], originating from the compositional fluctuation and other defects as shown clearly by HRTEM observation [22].

Fig. 2b shows the (004) diffraction line of the sample poled at 4 kV/cm. The profile is also symmetric, with a diffraction angle equal to the unpoled one (2$\theta$=87.50$^o$). But its FWHM ($\Delta 2\theta$=0.44$^o$) is much narrower than the unpoled sample. This means that the crystal form of this transition does not change, while the FE$_{LRO}$ (long range ordering rhombohedral ferroelectrics) can be perfectly induced in the poled sample with E$_{poling}$=4 kV/comfit is suggested that the effect of the random internal field can be overcome by an appropriate poling field. Furthermore the metastable FE$_{LRO}$ state can be locked even after the removal of the electric field in the room temperature. However upon thermal cycling a FE$_{LRO}$ to FE$_{SRO}$ phase transition of this sample occurres around 90 $^o$C. This phase transition can be regards as a "defrozen process" [23] and temperature 90 $^o$C is correlative to T$_f$ [15,22]. For the sample poled under E$_{poling}$=13 kV/cm, a quite different profile of the (004) diffraction line was observed, as shown in Fig. 2c. Although the plots of dielectric constant measurement Fig. 1b and 1c are almost the same, the peak of the (400) profile was shifted to 2$\theta$=86.60$^o$. Thus the lattice parameter $c$ of this form increases to 0.406nm, which is even larger than $c_m$=0.405nm of monoclinic phase as a transition phase around MPB between $FE_R$ phase and $FE_T$ phase [14]. Recently Noheda et al indicated that the tetragonal phase can be induced by an electric field larger than 10KV/cm applied along the [001] direction of PZN-4.5PT single crystal [24]. Therefore, in our case it is reasonable to suggest that a fraction of tetragonal ferroelectric state $FE_T$ has been induced by the stronger applying field, although the polarization rotation from <111> to <001> may not transform completely which can be presumed from the asymmetric diffraction peak shown in Fig.2c.

In addition to the XRD experiments, our explanation is also supported by the measurements of piezoelectric constant $d_{33}$. The value of $d_{33}$ for the sample poled at $E_{poling}$=4 kV/cm is higher than 1500 pC/N, but decreases to 850 pC/N for the sample poled at $E_{poling}$=13 kV/cm. The additional poling at 13kV/cm results in a phase transformation from the rhombohedral to the tetragonal phase, which reduces the contribution from the <111> polarization rotation.

In summary, these experiments indicate that the electric-field-induced phase transformation depends on the poling intensity for <001> oriented 0.7PMN-0.3PT single crystals. A weaker poling intensity ($E_{poling}$=4kV/cm ) can overcome the effect of random internal field to transform the phase from $FE_{SRO}$ to $FE_{LRO}$. But the rhombohedral ferroelectric to tetragonal ferroelectric phase transformation originated from <111> to <001> polarization rotation can only be induced by stronger poling intensity ($E_{poling}$=13 kV/cm). The $E_{poling}$=4 kV/cm poled sample shows a larger piezoelectric effect.

This work was supported by the National Natural Science Foundation of China Grant No.59995522. The authors would like to thank Dr. Newnham, Dr. Chen Jie and Dr. Fang Fei for their helpful discussion.

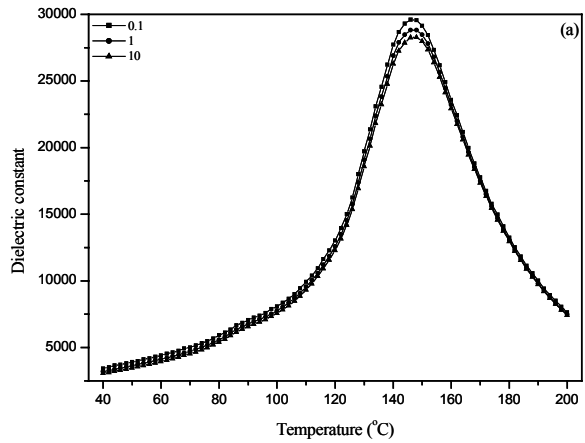

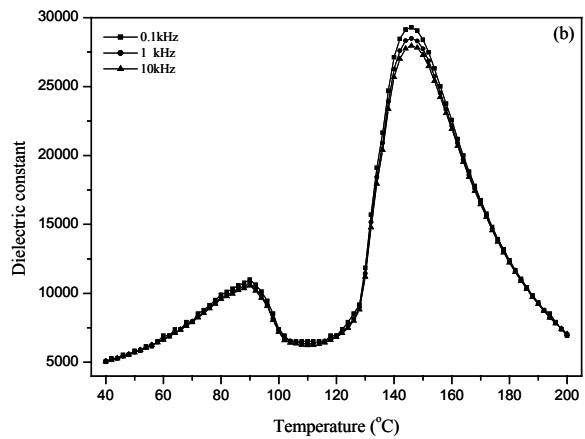

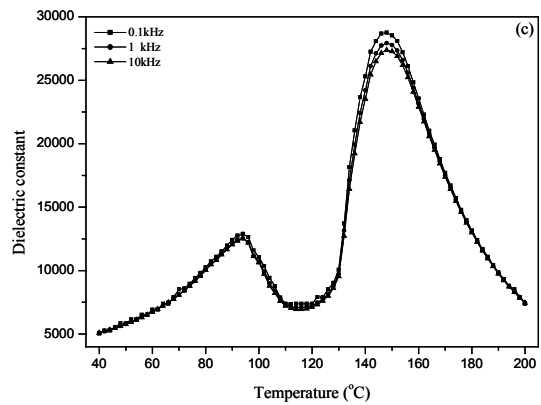

Fig.1 Shows the dielectric constant as functions of temperature for <001>oriented 0.7PMN-0.3PT single crystals in different frequency and poling condition (a) unpoled (b) $E_{poling}$=4 kV/cm (c) $E_{poling}$=13 kV/cm

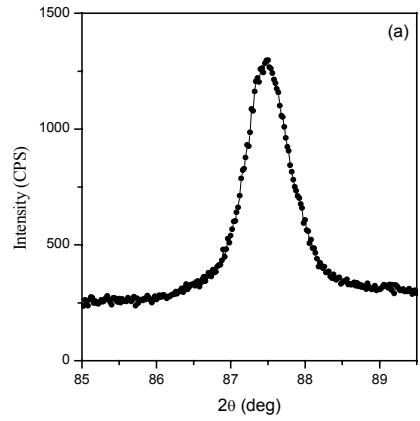

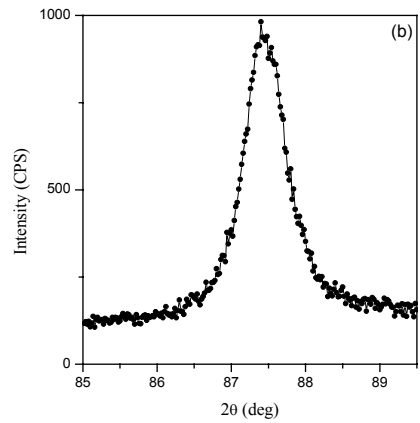

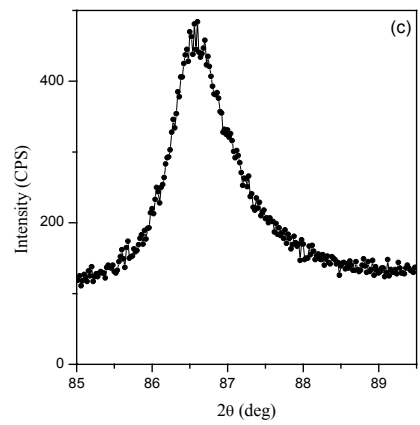

Fig.2 Shows the (004) X-ray diffraction lines of 0.7PMN-0.3PT single crystals (a) unpoled (b) poled at $E_{poling}$=4kV/cm and (c) poled at $E_{poling}$=13kV/cm. All were measured at room temperature.